# Using The Interactive Graphic Syllabus In The Teaching Of Economics

Seyyed Ali Zeytoon Nejad Moosavian, North Carolina State University, USA

**ABSTRACT**

*Syllabus is essentially a concise outline of a course of study, and conventionally a text document. In the past few decades, however, two novel variations of syllabus have emerged, namely "the Graphic Syllabus" and "the Interactive Syllabus". Each of these two variations of syllabus has its own special advantages. The present paper argues that there could be devised a new combined version of the two mentioned variations, called "the Interactive Graphic Syllabus", which can potentially bring us the advantages of both at the same time. Specifically, using a well-designed Interactive Graphic syllabus can bring about many advantages such as clarifying complex relationships; causing a better retention; needing less cognitive energy for interpretation; helping instructors identify any snags in their course organization; capability of being integrated easily into a course management system; appealing to many of learning styles and engaging students with different learning styles.*

*In addition to the introduction of the notion of the Interactive Graphic Syllabus, in order to put this idea into action in the context of economics, the present paper takes advantage of the visual "big picture" of intermediate macroeconomics that has already been proposed by Moosavian (2016a). The present paper describes a web-page that contains a web-based interactive graphic of the aforementioned macroeconomics visual "big picture". It is argued that this graphic can be used as a cyber-resource to technologically support the above-mentioned visual "big picture", which comprises twenty-seven interrelated macroeconomic diagrams, and gives some details on the types of their relationships. Furthermore, it provides numerous internet links to other relevant instructional resources offered by well-known universities, allowing the students to somehow zoom in the macroeconomics "big picture". Moreover, this web-page provides roughly one hundred links to short instructional videos. More interestingly, it responds interactively, if one hovers over a particular diagram, by highlighting the routes through which the final macroeconomic general equilibrium is affected by any change in that particular diagram. The interactive graphic went online on 10/30/2015 at the URL http://zeytoonnejad.com/macrobigpic.aspx and is still under minor modification.*

## 1. INTRODUCTION

Syllabus can be simply defined as a concise outline of a course of study. It also serves as students' introduction to their course, subject matter, and instructor. It usually contains the schedule of assignments, readings, and activities. It gives the students insight into the material and the instructor's approach to teaching that material. As Nilson (2010) puts it, it is "not only the road map for the term's foray into knowledge, but also travelogue to pique students' interest in the expedition and its leader." Nilson (2010) lists up 23 items as appropriate items that should be included in any syllabus.[1] Grunert (1997) proposes a learning-centered syllabus that incorporates considerably more information than the basic syllabus routines suggest. Some, but not all, of these additional items are as follows: instruction on how to use the syllabus, recommendations on how to study for the course along with learning tools and aids, course content information, etc. This expansion of the notion of syllabus and defining the additional roles for a syllabus to play necessitates us to define new forms and novel variations of syllabus.

[1] Nilson (2009) goes beyond that and lists 44 items as the basic information that should appear in a syllabus. Through the years, the typical syllabus has grown in size, from a short one- to two-page schedule of course topics, assignments, and tests to a five- to ten-page list of information required by institutions, some accrediting agencies, departments, and students themselves. When a syllabus becomes this long, a natural question arises on how to get students to read our syllabus. To answer this question you can read Nilson (2010) that provides 4 options on how to get them to read your syllabus.

Conventionally, a syllabus is a text document. In the past few decades, however, two novel variations of syllabus have been introduced, namely "the Graphic Syllabus" and "the Interactive Syllabus". The Graphic Syllabus, which has been primarily introduced by Nilson (2009), is defined as "a flowchart or diagram that displays the sequencing and organization of major course topics through the semester." The Interactive Syllabus, which has been primarily introduced by Richards (2003), is defined as "a web-based tool that offers rich, robust media that appeals to students' different learning styles, and normally includes not only texts, but also full length texts of online books, publishers' online course materials, images, audio, and video." Each of these two variations of syllabus has its own special advantages, which will be discussed in greater detail in the next section. The present paper argues that there could be devised a new combined version of the two mentioned variations, called "the Interactive Graphic Syllabus", which can potentially bring us the advantages of both at the same time. The main purpose of the present paper is to introduce this third variation in a general sense and provide a practical example for the case of intermediate macroeconomics to show how easily one can put the idea into practice in the context of economics.

In order to put the notion of the Interactive Graphic Syllabus into action in the context of economics, the present paper takes advantage of the visual "big picture" of intermediate macroeconomics that has already been proposed by Moosavian (2016a). As Moosavian (2016a) explains, macroeconomics textbooks often introduce macroeconomic diagrams and models in separate sections, obscuring the understanding of the relationships among those diagrams and models. As mentioned above, Moosavian (2016a) has designed a so-called "big picture" for macroeconomics to highlight those relationships. Moosavian (2016c) elaborates the components of this visual in greater detail. In brief, the aforementioned visual "big picture" is utilized to provide a practical example of how one can put the idea of the Interactive Graphic Syllabus into practice for the case of intermediate macroeconomics.

The present paper also describes a web-page that contains a web-based interactive graphic of the macroeconomics "big picture" introduced by Moosavian (2016a). This graphic can be applied as a cyber-resource to technologically support the above-mentioned visual "big picture", which comprises twenty-seven interrelated macroeconomic diagrams, and gives some details on the types of their relationships. Furthermore, it provides numerous internet links to other relevant instructional resources offered by well-known universities, allowing the students to somehow zoom in the macroeconomics "big picture". Moreover, this web-page provides roughly one hundred links to short instructional videos. More interestingly, it responds interactively, if one hovers over a particular diagram, by highlighting the routes through which the final macroeconomic general equilibrium is affected by any change in that particular diagram. There are many more built-in features, capabilities, and menus offered by this web-page which will be introduced in section 3.

Concerning the novelty of the ideas presented in the present paper, it should be noted that this paper is the first attempt to introduce the idea of "the Interactive Graphic Syllabus" in general. Consequently, it is also the first attempt made to discuss such an idea in economics and intermediate macroeconomics in particular. As a result, the objective of the paper is to introduce the Interactive Graphic Syllabus and propose an Interactive Graphic Syllabus for the course of intermediate macroeconomics as a practical example.

The remainder of the paper has been organized such that a short literature review on the two main modern variations of syllabus is undertaken in the next section. In section 2.1, the Graphic Syllabus is reviewed, and subsequently, in section 2.2, the Interactive Syllabus is reviewed. In section 3, the main discussion of the paper is presented. In particular, in section 3.1, the notion of the Interactive Graphic Syllabus as a third combined variation of the two apparently separate variations which have been introduced already. Afterwards, section 3.2 puts the idea into practice for the case of intermediate macroeconomics. Next, some tips on designing an Interactive Graphic Syllabus are offered in section 3.3. Finally, a conclusion on the whole discussion will be drawn in section 4.

## 2. LITERATURE REVIEW: MODERN VARIATIONS OF SYLLABUS

In this section, two modern variations of syllabus are discussed primarily in the form of a literature review. Finally, this discussion will be employed to build up a framework for a third novel variation which is to be introduced by the present paper in section 3.

### 2.1 The Graphic Syllabus

Traditionally and by default, a syllabus is thought to be a text document. However, over the years, several novel types of syllabus have emerged, one of which is the Graphic Syllabus. A graphic syllabus is a visual tool that is designed to communicate a course to students more effectively. The idea of the Graphic Syllabus has been primarily introduced by Nilson (2009) in her textbook called "The graphic syllabus and the outcome map: Communicating your course". According to Nilson (2009), the Graphic Syllabus can be defined as "a flowchart or diagram that displays the sequencing and organization of major course topics through the semester." She goes on to further explain that "it uses spatial concept arrangement, connecting lines, arrows, and sometimes numbers to show the logical, temporal progression of the course through topics within the subject matter." In this sense, a graphic syllabus is the skeletal structure of a course, and in a deeper sense, as Nilson (2009) describes the Graphic Syllabus, it is a piece of scholarship, one that brings the scholarship of integration to the scholarship of teaching (Boyer, 1990). She also believes that for any given course, "our syllabi display our conception of how a field or sub-field is organized - or should be organized for the purpose of communicating it – and how students can best master its knowledge and skills."

In supporting the idea of the Graphic Syllabus, Nilson (2009) explains the pedagogical power of graphics by referring to dual-coding theory and the visual argument theory. She also gives some evidence on the power and potential of graphics in increasing the efficiency of learning and retention, as well as reaching visual, global and intuitive learning styles. However, further elaboration of these pieces of evidence is beyond the scope of the present paper.

Undoubtedly, using a well-designed graphic syllabus can bring about many advantages. These benefits could include clarifying complex relationships; giving an additional level of understanding (by giving the "big picture" of the course content); revealing why the instructor has organized the course the way she did; causing a better retention[2]; needing less cognitive energy for interpreting a graphic; better communicating than text across cultural and language barriers; showcasing alternative ways of taking notes, outlining papers, and organizing concepts; and helping instructors identify any snags in their course organization.

Nilson (2009) also explains why it is important to design a graphic syllabus, how it facilitates learning from a theoretical point of view, and how to design a graphic syllabus to reflect the structure of a course. She also provides additional annotated examples of graphic syllabi, and finally she concludes with information on available software for designing a graphic syllabus.[3]

### 2.2 The Interactive Syllabus

A second modern type of syllabus that has lately emerged is the Interactive Syllabus. The idea of the Interactive Syllabus has initially been introduced and discussed by Richards (2003). In her paper entitled "The Interactive Syllabus: A resource-based, constructivist approach to learning", she explains that "the Interactive Syllabus is a tool fashioned along constructivist principles that facilitates the implementation of broader constructivist strategies in an internet-delivered course site. Hence, it is a constructivist approach that attempts to solve some fundamental problems in course delivery and design[4].” As she puts it, by the term “interactive” she refers to an environment in which concepts are presented in various ways resulting in multiple and adaptive interpretations necessary for knowledge acquisition. She argues that although there is nothing inherently “sizzling” about the syllabus, but by employing the notion of the Interactive Syllabus we can make the syllabus “a robust, content-rich learning environment that initiates the student into a constructivist learning space.”

[2] The Graphic Syllabus is essentially a visual representation. Research has shown that students are more likely to comprehend and remember the materials that they receive both verbally and visually. A good visual is actually hard to forget.
[3] The list of this software programs have been reported in appendix 6.

She also states that the necessity of developing interactive syllabi is very often confused with “putting content” into a course site. She divides the items to be included into a syllabus into two broad categories, namely the “prescriptive items” that prescribe the students on the specifics of the course, and the “descriptive features” that outline the topics to be studied. Examples of the prescriptive items include “course objectives”, “course requirements”, and “grading policies.” Examples of the descriptive features include the structure and logical order and sequence of the course material. The Interactive Syllabus attempts to employ these descriptive features and to generate a robust learning space in an open format that allows self-determined investigation. In this way, the Interactive Syllabus becomes the initial source of course knowledge that will permit students to enter into fruitful and deep study explorations.

According to Richards (2003), the main idea behind the Interactive Syllabus is to “populate the syllabus with rich, robust media that will appeal to students’ different learning styles.” The media can be of different forms such as texts, full length texts of online books, publishers’ online course materials, images, audio, and video.

There is no doubt that using a well-designed interactive syllabus can bring about several advantages. These advantages could include offering learners a pedagogically rich portal to course materials that form the basis of the course investigation; capability of being integrated easily into a course management system, or part of an online course with an open architecture; exploiting the descriptive features of a course and providing a robust, content-rich learning environment that initiates students into the creative world of a constructivist learning space in an open format that permits self-determined investigation; appealing to many of learning styles in its use of materials; and providing a relatively simple way to engage students with different learning styles (Richards, 2003).

Although there have been published a few excellent books and papers on some of the new variations of syllabus over the past few decades (see, for example, Nilson (2010), Nilson (2009), and Richards (2003)), there has never been any paper putting the two separately-discussed ideas of the Graphic Syllabus and the Interactive Syllabus together, which could potentially benefit from the advantages brought about by both. In this sense, the present paper is to fill this gap in the literature. Overall, this paper makes three contributions. The first contribution of the paper is to review the features of the two already-existing modern variations of syllabus, which was already done in section 2. The second contribution is to introduce a third combined variation, namely the Interactive Graphic Syllabus, which is to be done in section 3.1. Finally, the third contribution is to provide a practical example of how to put the notion of the Interactive Graphic Syllabus into practice in the context of economics and in particular for the case of intermediate macroeconomics, which is to be carried out in section 3.2.

## 3. MAIN DISCUSSION

In this section, a third variation of modern syllabus, called the Interactive Graphic Syllabus, is introduced first. After that, the introduced idea is put into practice by presenting a practical example for the course of intermediate macroeconomics.

### 3.1. A Third Combined Variation: The Interactive Graphic Syllabus:

Each of the two modern variations of syllabus, which were discussed in section 2, possesses its own advantages. In order to gain the advantages of both of the aforementioned variations at the same time, a wise idea and a good practice is to put together these two types, and come up with a third combined variation, called the Interactive Graphic Syllabus, in which essentially a wisely-designed graphics syllabus is placed in an online interactive environment. A wisely-designed graphic syllabus is ideally one through which the structural framework of the course being offered is visually communicated. An online interactive environment is preferably one in which the conceptual components of the structure are presented through various forms (including texts, graphics, short audios, short videos, etc.) in order to serve and appeal to a wide variety of learning styles[5].

[4] “Constructivism is a learning theory found in psychology which explains how people might acquire knowledge and learn. It therefore has direct application to education. The theory suggests that humans construct knowledge and meaning from their experiences. Constructivism is not a specific pedagogy. Piaget's theory of Constructivist learning has had wide ranging impact on learning theories and teaching methods in education and is an underlying theme of many education reform movements” (The University of Sydney, 2016). You can see more at: http://sydney.edu.au/education_social_work/learning_teaching/ict/theory/constructivism.shtml#sthash.HvYwX8Yj.dpuf

Therefore, the Interactive Graphic Syllabus can be defined briefly as “the visual ‘big picture’[6] of a course which is accessible in an online environment in a way that students can easily interact with the elements of the course structure.” More specifically, this visual “big picture” should ideally be the “big picture” of the related field (or sub-field) in the form of a flowchart or diagram that displays the sequential and logical order of the materials in that particular field[7]. In addition, it must be accessible in an online space in such a way that students can interact with the elements of the course structure and somehow zoom in the big picture of it if they opt to do so. Such an electronic document will depict the whole organization of major elements of the course to be discussed throughout the semester.

No matter whether the course that you are teaching is offered in a traditional, or well-facilitated, or hybrid (blended), or fully-online fashion, the Interactive Graphic Syllabus can be effectively utilized and become a vital component of a successful virtual learning environment. The appearance and theme of the Interactive Graphic Syllabus remains discretionary with every instructor. It is important that the instructor search for web-based resources used to put together the Interactive Syllabus[8].

The resources collected from the Internet must be appropriate in terms of subjects, and also target various learning styles. For example, global learners will usually benefit from the Interactive Graphic Syllabus by looking at the comprehensiveness of the structure which is provided by the visual “big picture” of the course or ideally the field. Sequential learners will routinely benefit from the sequential and possibly logical orders of the components included in the visual “big picture”. Visual learners will commonly benefit from pictures, diagrams, maps, videos, and animations. Auditory learners will appreciate audio files. Kinesthetic learners will take advantage of materials with controls that allow them to regulate the way in which they interact with materials. A correctly-designed interactive graphic syllabus should serve and appeal to all of the above-mentioned learning styles in its use of materials[9].

Finally, it is important to remind and recognize that the Interactive Graphic Syllabus is not a tool by which you can achieve all the learning objectives of a course. Lectures, lecture notes, face-to-face discussions, texts, papers, assignments, problem-solving, quizzes, exams, projects, and many other forms of teaching and learning activities and tools will help us achieve the various learning objectives of the course. The Interactive Graphic Syllabus is, however, an excellent tool to communicate the structure of the course of study in a visual and effective way, which is not possible to be done so easily and effectively through the other teaching tools mentioned above. As Richards

[5] It is important to clearly differentiate between the idea of the Interactive Syllabus and the Interactive Graphic Syllabus. They are different in that although in the Interactive Syllabus there are multi-media components, but in its definition, there is no mention of “the visual big picture” of the course or field, which is the main component of the Graphic Syllabus. In fact, the Interactive Graphic Syllabus included the notion of the visual “big picture” into the Interactive Syllabus, which results in a new combined variation of syllabus called the Interactive Graphic Syllabus.

[6] In his paper entitled “Teaching economics and providing visual ‘big picture’ ”, Moosavian (2016a) discusses the importance, necessity, and advantages of providing visual “big pictures” in the teaching of economics. Thus, for further information on this topic, you might want to see the mentioned paper.

[7] A natural question that may arise here is to ask what if an instructor is not supposed to teach all the elements of the field or sub-field associated with that particular course. A reasonable answer would be as follows: Even in this case, by providing the structure of the field, instructors will help their students know what parts of the field they are supposed to learn about and what parts they are NOT supposed to learn about in that particular semester, although it is part of the field. In general, knowing what we do not know yet can be helpful in achieving a comprehensive understanding of the “big picture” of the field.

[8] According to Richards (2003), as a general practice, “students should not be assigned the task of finding appropriate links for course topics. By and large, they do not possess the information literacy skills required to undertake this activity.”

[9] Needless to say, besides the advantages mentioned here, the Interactive Graphic Syllabus essentially inherits and naturally nests those of both the Interactive Syllabus as well as the Graphic Syllabus that I listed up in the previous section after introducing either one. For the sake of space economy and compactness of the paper, I fail to re-list those advantages which I listed in two separate paragraphs in the previous section.

(2003) describes a good interactive syllabus, a good interactive graphic syllabus must be the "locus amoenus"[10] for cognitive scaffolding and reflective thinking. According to Richards (2003), "cognitive scaffolding means that a student needs to visit a learning space more than once in order to construct meaning. Each visit builds upon prior knowledge as well as new experiences gained from explorations. In between each visit, there must be time for reflective thinking." Therefore, a well-designed interactive graphic syllabus must furnish students with such a pleasant learning environment.

### 3.2. Putting the Idea into Practice: Case of Intermediate Macroeconomics

Up to this point in this paper, I have built up the idea of the Interactive Graphic Syllabus in general terms. Now, in order to put this idea in practical terms and clarify how one can put the notion into action in the context of economics, a fine example of the Interactive Graphic Syllabus for the course of intermediate macroeconomics is provided in this subsection. To this end, the visual "big picture" of intermediate macroeconomics developed by Moosavian (2016a)[11] is borrowed, and then the graphic is used to develop a typical example of the Interactive Graphic Syllabus.

The visual "big picture" introduced by Moosavian (2016a), which is also presented in appendix 1, contains 27 macroeconomic diagrams which are commonly discussed in the intermediate macroeconomics courses. Of these, 13 diagrams help users understand how to derive aggregate supply, and 12 diagrams help users to see how to derive aggregate demand. Appendix 2 describes the visual "big picture" of intermediate macroeconomics in terms of different markets and main emphases of macroeconomic analyses conducted by the two traditional mainstream economics schools of thought. An alphabetical list of the diagrams and models included in this figure is provided in appendix 3. In appendix 4, a list of symbols, signs, and notations employed in the figure is presented.[12] This visual "big picture" will constitute the graphic part of the Interactive Graphic Syllabus, which visually communicate the structural framework of intermediate macroeconomics. Additionally, the figure has been placed in an online interactive environment, in which the conceptual components of the structure are presented through various forms of media in order to serve and appeal to a wide variety of learning styles. Therefore, this mixture is now indeed a visual 'big picture' of intermediate macroeconomics which is accessible in an online environment in a way that students can easily interact with the elements of the course structure. Figure 1 demonstrates a snapshot of the web-page on which the Interactive Graphic Syllabus being introduced is accessible. The interactive graphic went online on 10/30/2015 at the URL http://zeytoonnejad.com/macrobigpic.aspx and is still under minor modification. This web-page has been designed such that it works properly on almost all popular internet browsers, such as Firefox, Internet Explorer, Chrome, and Safari.

---

[10]. A Latin expression for "pleasant place" referring to an idealized place of safety or comfort

[11]. Macroeconomics textbooks often introduce macroeconomic diagrams and models in separate sections, obscuring interrelationships among those diagrams and models. Moosavian (2016a) has designed a helpful visual "big picture" for intermediate macroeconomics that highlights these interrelationships.

[12]. For further information on how this visual "big picture" has been designed and should be used, you might want to see Moosavian (2016c).

**Figure 1.** A snapshot of the web-page of the Interactive Graphic Syllabus[13]

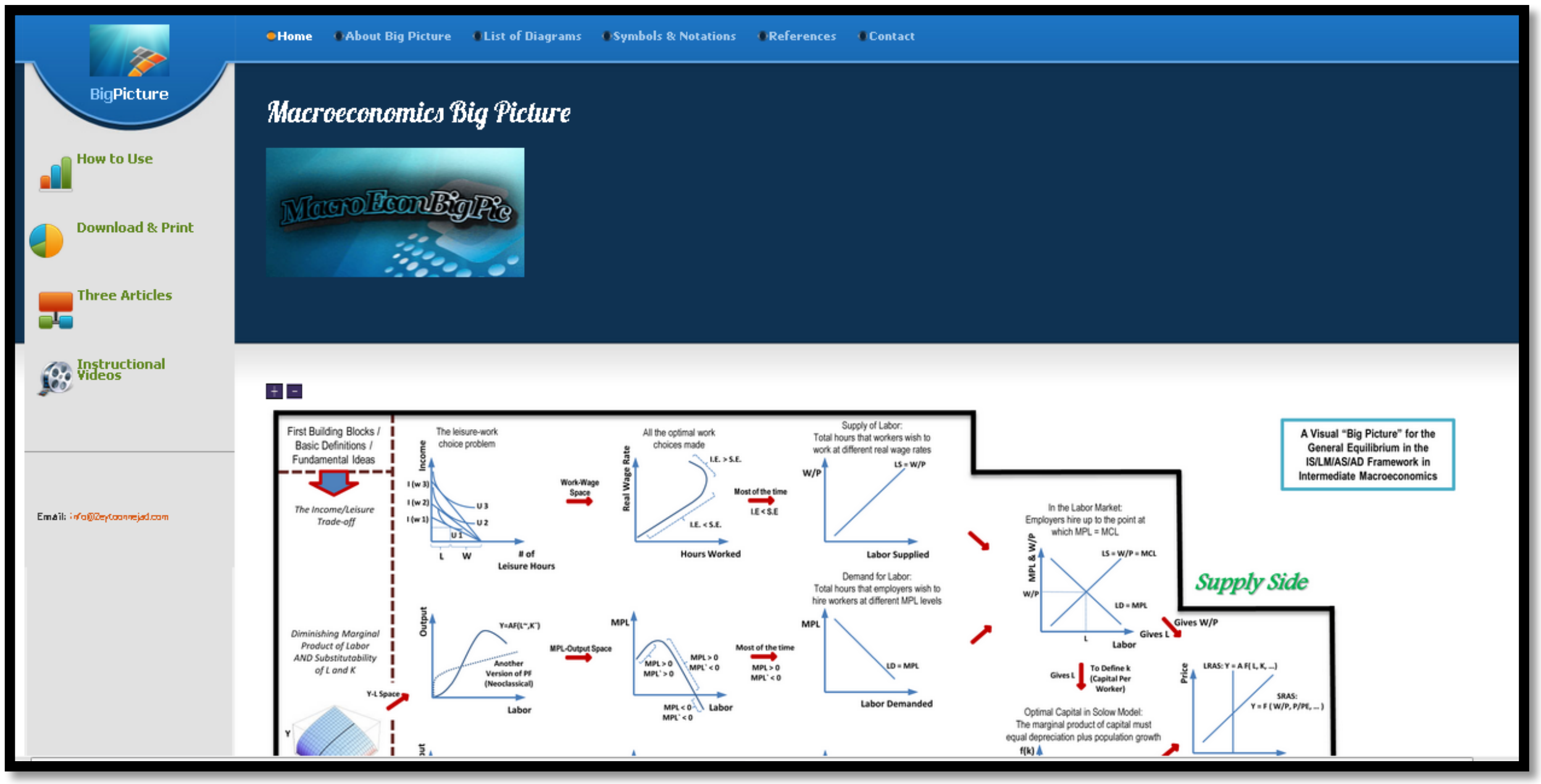

This visual "big picture" is an ideal one since it depicts the "big picture" of the related field (as opposed to just that of the course) in the form of interrelated diagrams that display the sequential and logical order of the materials. It is also accessible in an online space in such a way that students can interact with the components of the course structure and somehow zoom in the "big picture" if they wish to. This electronic document demonstrates the whole organization of the major components of intermediate macroeconomics that are usually discussed during the semester.

There are numerous built-in features, capabilities, and menus that this web-page offers. Among those, there are four main features that are worthwhile to be specifically described here. Firstly, this interactive web-page explains the types of the relationships among the individual diagrams by some brief notes on the red arrows.[14] Secondly, it provides multiple links for each of the individual diagrams, which describe, by texts and graphics, how each diagram is derived from another. These texts have been selected from available online resources provided by various well-known universities. These links help users gain more detailed information on each of the diagrams. Another capability offered by this interactive web-page is that it demonstrates in what ways a change in a diagram or model can have an impact on the final macroeconomic general equilibrium. More specifically, this web-page responds interactively, if one hovers over a particular diagram for five seconds, by highlighting the routes through which the final macroeconomic general equilibrium is affected by any changes in that particular diagram. Figure 2 shows this feature for the case of leisure-work diagram. Moreover, this web-page provides about 5 to 10 links to easy-to-digest instructional videos that describe each of the individual diagrams and their relationships with the others related to that. In total, it provides about one hundred links to appropriate short instructional videos.

[13] As shown in figure 1, on top and the right of the visual "big picture", there are numerous menus that provide different capabilities, but their explanations are beyond the scope of the present paper. In addition, to see how this visual can be applied, you can see Moosavian (2016c).
[14] As Moosavian (2016a) explains, these relationships are in fact of three types, namely "derivative relationship", "common-part relationship", and "perspective relationship". For more information on these types, you can see Moosavian (2016c).

**Figure 2.** A snapshot of the response of the web-page to hovering over a diagram

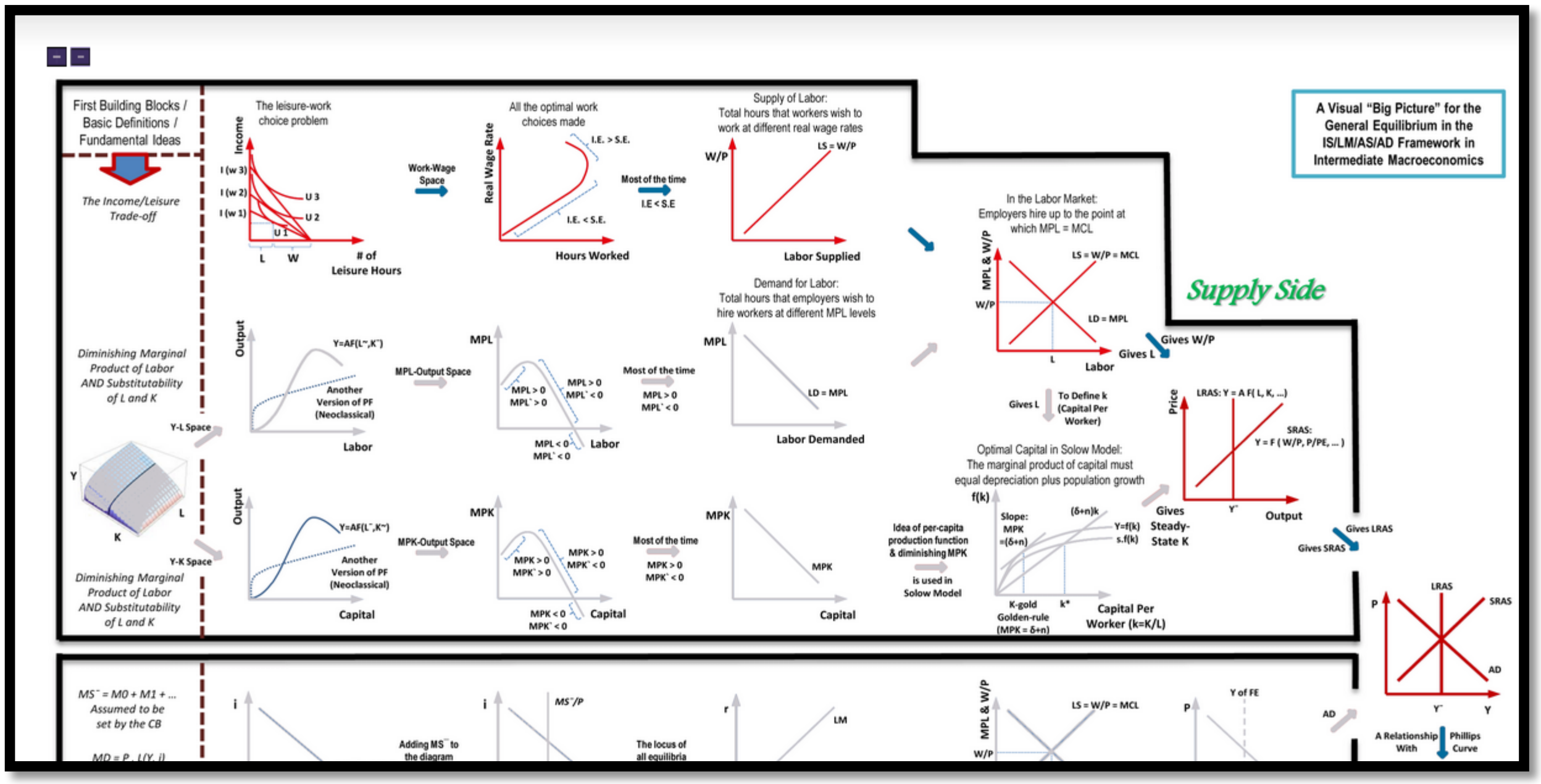

This interactive graphic is designed for users who are interested in both having a big picture of macroeconomics and gaining specific information about the individual macroeconomic diagrams and models, and as such, it is a great achievement. With this innovation, there is no longer a trade-off between getting the big picture and digging into the material. That is to say, one can do both at the same time, and see where in macroeconomics they are digging into. It can be applied as a cyber-resource to technologically support the described visual "big picture" and the Interactive Graphic Syllabus. Appendix 5 provides the graphic part of the paper version of the Interactive Graphic Syllabus being discussed in this paper, which uses the visual "big picture" and the interactive web-page described above as a complementary tool. In the paper version, the order of chapters has also been shown by putting the numbers of the chapters in red on each of the diagrams corresponding to that particular chapter. This task has been done in accordance primarily with a macroeconomics textbook called "macroeconomics" by Abel and Bernanke (2008).

The Interactive Graphic Syllabus presented here applied the visual "big picture" of intermediate macroeconomics as an example. To see two additional examples of visual "big pictures" in the context of economics you can see Naumenko and Moosavian (2016) and Moosavian (2016b) and Moosavian (2016e). The first paper presents the visual "big picture" of production theory for the course of advanced microeconomics. The last two papers present the visual "big picture" of consumer theory for the course of advanced microeconomics (Ph.D. core course). These two figures are two additional examples that can be used for designing the Interactive Graphic Syllabus for graduate (advanced) microeconomics courses.

To wrap up this section, there are two additional points that are noteworthy. First, it is important to note that the graphic part of the Interactive Graphic Syllabus does not have to show the structure of the 'field', although it will be ideal if it does so, for the reason mentioned in footnote 8. However, at a minimum, it should show the structure of the 'course'. Second, for any syllabus, text and graphic should be regarded as complements, not substitutes. This is because a major capability of text is to explain details well while the primary task of a graphic is to depict the structure of a course. More generally and broadly speaking, text can be used to communicate "prescriptive features" (e.g., attendance policy, and grading policy) of a course while graphics should be used to communicate "descriptive features" (e.g., the logical order of the material) of the course.

In the next section, a few tips on designing such an interactive, web-based graphic will be offered.

## 4. SOME TIPS ON DESIGNING YOUR INTERACTIVE GRAPHIC SYLLABUS

In order for a visual "big picture" to be considered well-designed, there are multiple handy tips that should be followed. Here is a short list of useful tips on designing good Interactive Graphic Syllabi.

- *Remain consistent with your notation.* Do not forget the fact that one of the missions of a visual "big picture" is to help students remove confusions and misunderstandings, so be cautious not to make new confusions and misunderstandings by using inconsistent, misleading notations.
- *Avoid going into much detail.* Do not put too much detail on a visual "big picture". It is in fact against the main philosophy of providing a visual big picture. An overly detailed "big picture" could make matters worse.
- *Design on the basis of the learning objectives of the courses*. As suggested by Davenport et al. (2008), considering the learning goals, the design of visual representations, and the learners' abilities when designing visual presentations could make your syllabi more effective[15].
- *Put necessary notes that help avoid confusions*. You can put some little notes on your visual "big picture" as well if you find them necessary or helpful to avoid confusion. Nevertheless, adhere to the principle that they need to be short enough.
- *Take advantage of color-coding if necessary.* Utilize two or more colors as a means of distinguishing among different levels of importance associated with major items and minor items, or as a means of indicating relationships among the elements of your visual "big picture".
- *Sometimes, even apparently minor things matter.* There are some cases where even apparently minor characteristics of an element such as its thickness or color could be important and informative. For instance, the thickness of a line could matter and, say, be an indication of the importance of that particular line compared to the other lines, if you design and define it to mean so.
- *Tidiness matters*: A visual "big picture" needs to be arranged meticulously and neatly. An untidy visual "big picture" only makes matters worse, and does not help students build up their desired well-designed mental framework of the course or that of the field.
- *The simplest version of a multi-version diagram is often preferred to its more complex versions*: Although there are sometimes multiple versions of the same diagram (e.g. Solow model with and without considering population growth and/or technological progress), but we usually focus on the simplest version available, since the main task of this tool - the visual "big picture" - is to help students understand the underlying linkages, not to help them recognize the different versions of an essentially the same idea.
- *A Trade-off between Holding Simplicity and Elaborating Complexity*: Just as many other circumstances in the real world, so we, as economics instructors, have to finally make a choice of the optimum combination of holding simplicity and elaborating complexity. In other words, on the one hand, holding simplicity helps us stick to the philosophy of a visual "big picture"; on the other hand, elaborating complexities further may help students avoid confusions; however, this task is often better to be done primarily through the use of other teaching and learning tools such as lectures, textbook, etc., except for some rare points that are absolutely necessary to be cleared up in the visual "big picture" as well. It is also important to note that oversimplification of a "big picture" cannot help economics students sufficiently to get the whole picture and get rid of the potential confusions. This is because such a "big picture" would ignore many major parts of the framework of the course which do exist in reality. As a result, try your best to make your graphic as comprehensive as possible (in terms of structure, and not details).
- *You can use different software programs to design the graphic part:* To design the graphic part of your Interactive Graphic Syllabus, you can use many different software programs. One of the most convenient software programs that can be used for this purpose is simply Microsoft PowerPoint. However, if you want to use the computer software that have been designed specifically for this purpose, you can refer to the list of the programs reported in table B.1 presented in Nilson (2009), which is also presented in appendix 6 of the present paper.

[15] In order to obtain more information on how to design diagrams with desirable features from a cognitive point of view, you can read the article entitled "When do diagrams enhance learning? A framework for designing relevant representations," written by Davenport et al. (2008).

- *You can use different ways to design the interactive part:* First of all, it is important to clarify that the appearance and theme of the Interactive Graphic Syllabus remains discretionary with every instructor. Secondly, to design the interactive part of your Interactive Graphic Syllabus, you have different alternative ways to take advantage of. Of course, the easiest way to get that part done is to hire a professional for this purpose. Then, as the content specialist, you as the instructor give the professional your designed visual "big picture", preferred Web-based resources, and selected architecture to connect the graphic with the resources to be all included in your interactive graphic syllabus. However, if you intend to do and insist on doing the interactive part by yourself, then after finding online materials, you store the resources in a Word document, and then use an HTML editor (such as Adobe Dreamweaver) to construct the page. If you are interested in more sophisticated software you can use Flash for instance.
- *Link directly to the file of interest*: When adding text and video files which you have found on the Web, it is important to link directly to the text or media files themselves, and not to the pages that host them for now.
- *Ask for help or hire a professional for the parts you have no clues on how to get them done:* In total, designing an interactive graphic syllabus involves two essential phases: The design part for the graphic and the Web-related part. Get help with either or both if needed.

These are indeed the tips you can take advantage of in practice when designing your own visual "big picture", and this list can go on and on, but I stop it here to keep things short. The next section of the paper will draw a conclusion on the whole discussion.

## 5. CONCLUSION

Syllabus is essentially a concise outline of a course of study, and conventionally a text document. In the past few decades, however, two novel variations of syllabus have emerged, namely "the Graphic Syllabus" and "the Interactive Syllabus". Each of these two variations of syllabus has its own special advantages, which were discussed in detail throughout the paper. The present paper argued that there could be devised a new combined version of the two mentioned variations, called "the Interactive Graphic Syllabus", which can potentially bring us the advantages of both at the same time. Specifically, using a well-designed Interactive Graphic syllabus can bring about many advantages, such as clarifying complex relationships; causing a better retention; needing less cognitive energy for interpretation; better communicating than text across cultural and language barriers; helping instructors identify any snags in their course organization; offering learners a pedagogically rich portal to course materials that form the basis of the course investigation; capability of being integrated easily into a course management system, or part of an online course with an open architecture; appealing to many of learning styles in its use of materials and providing a relatively simple way to engage students with different learning styles.

In addition to the introduction of the notion of the Interactive Graphic Syllabus, in order to put this idea into action in the context of economics, the present paper took advantage of the visual "big picture" of intermediate macroeconomics that had already been proposed by Moosavian (2016a). The present paper also described a web-page that contains a web-based interactive graphic of the aforementioned macroeconomics visual "big picture". It is argued that this graphic can be used as a cyber-resource to technologically support the above-mentioned visual "big picture", which comprises twenty-seven interrelated macroeconomic diagrams, and gives some details on the types of their relationships. Furthermore, it provides numerous internet links to other relevant instructional resources offered by well-known universities, allowing the students to somehow zoom in the macroeconomics "big picture". Moreover, this web-page provides roughly one hundred links to short instructional videos. More interestingly, it responds interactively, if one hovers over a particular diagram, by highlighting the routes through which the final macroeconomic general equilibrium is affected by any change in that particular diagram. The interactive graphic went online on 10/30/2015 at the URL http://zeytoonnejad.com/macrobigpic.aspx and is still under minor modification. This web-page can serve different types of learners such as visual, auditory, kinesthetic, sequential, and global learners by providing them with texts, short audio-videos, interactive features, ordered diagrams, as well as a holistic, visual "big picture", respectively.

Overall, this paper made three contributions which are as follows: reviewing the features of the two already-existing modern variations of syllabus, introducing a third combined variation, namely the Interactive Graphic Syllabus, and providing a practical example of how to put the notion of the Interactive Graphic Syllabus into practice in the context of economics and in particular for the case of intermediate macroeconomics. The paper also presented a short list of useful tips on how to design good Interactive Graphic Syllabi.

## AUTHOR BIOGRAPHY

**Seyyed Ali Zeytoon Nejad Moosavian** is a Graduate Teaching Assistant and a fourth-year Ph.D. student in Economics in the Department of Economics at North Carolina State University. He received his Bachelor's degree in Mining Engineering, his first Master's degree in Economics (Applied Track) from Azad University of Iran, and his second Master's Degree in Economics (Theoretical Track) from North Carolina State University. His current teaching interests include Microeconomics, Macroeconomics, and Engineering Economy. His current research is focused on the Economics of Risk and Insurance as well as Economics Education and Pedagogy. He is also a member of the editorial boards of a number of scholarly journals such as *Journal of Business and Economic Research*, *Applied Economics and Finance*, and *International Journal of Economics and Finance*. He has published several scholarly research papers in the area of Economics Education and Pedagogy. He has primarily focused on the utilization of conceptual visualization techniques to decode theoretical complexities of various economic theories. His work in the area of Economics Education is both innovative and high-tech. Email: szeytoo@ncsu.edu

## REFERENCES

Abel, A. B., Bernanke, B. S., & Croushore, D. (2008). Macroeconomics (6th edn). Boston: Pearson Education, IncAdanu K (2005) A cross-province comparison of Okuns coefficient for Canada. Appl Econ, 37(5), 561570Aghion.

Boyer, E. L., & Reconsidered, S. (1990). Scholarship Reconsidered: Priorities of the Professoriate. Carnegie Foundation for the Advancement of Teaching, Princeton, NJ.

Davenport, J. L., Yaron, D., Klahr, D., & Koedinger, K. (2008, June). When do diagrams enhance learning? A framework for designing relevant representations. In Proceedings of the 8th international conference on International conference for the learning sciences-Volume 1 (pp. 191-198). International Society of the Learning Sciences.

Moosavian, S. A. Z. N. (2016a). Teaching Economics and Providing Visual "Big Pictures". *Journal of Economics and Political Economy*, *3*(1), 119-133. http://dx.doi.org/10.1453/jepe.v3i1.631

Moosavian, S. A. Z. N. (2016b). A Comprehensive Visual "Wheel of Duality" in Consumer Theory. *International Advances in Economic Research*, 1-2. http:/dx.doi.org/ 10.1007/s11294-016-9586-8

Moosavian, S. A. Z. N. (2016c). The Visual "Big Picture" of Intermediate Macroeconomics: A Pedagogical Tool to Teach Intermediate Macroeconomics. International Journal of Economics and Finance, 8(9), 234. http://dx.doi.org/10.5539/ijef.v8n9p234

Moosavian, S. A. Z. N. (2016d). Teaching Economics and Providing Visual" Big Pictures". *arXiv preprint arXiv:1601.01771*. Retrieved from http://arxiv.org/abs/1601.01771

Moosavian, S. A. Z. N. (2016e). The Visual Decoding of the "Wheel of Duality" in Consumer Theory in Modern Microeconomics: An Instructional Tool Usable in Advanced Microeconomics to Turn "Pain" into "Joy". *Applied Economics And Finance*, 3(3), 288-304. http://dx.doi.org/10.11114/aef.v3i3.1718

Naumenko, A., & Moosavian, S.A.Z.N. (2017). A holistic visual "wheel of relationships" in production theory, a poster presented at the American Economic Association (AEA) poster session at the 2017 ASSA/AEA Meeting, Chicago, IL, USA.

Naumenko, A., & Moosavian, S. A. Z. N. (2016). Clarifying theoretical intricacies through the use of conceptual visualization: Case of production theory in advanced microeconomics. Applied Economics and Finance, 3(4), 103-122. https://doi.org/10.11114/aef.v3i4.1781

Nilson, L. B. (2009). The Graphic Syllabus and the outcomes map: Communicating your course (Vol. 137). John Wiley & Sons.

Nilson, L. B. (2010). Teaching at its best: A research-based resource for college instructors. John Wiley & Sons.

Richards, S. L. (2003). The Interactive Syllabus: A resource-based, constructivist approach to learning. The Technology Source.

The University of Sydney (2016). Teaching with ICT: Theory, Practice, and Examples – Constructivism, Sydney, Australia. Available at: http://sydney.edu.au/education_social_work/learning_teaching/ict/theory/constructivism.shtml#sthash.HvYwX8Yj.dpuf

Zeytoon Nejad Moosavian S.A. (2016, June). Employing Technology in Providing an Interactive, Visual "Big Picture" for Macroeconomics: A Major Step Forward towards the Web-Based, Interactive, and Graphic Syllabus, Paper presented at the Sixth Annual American Economic Association (AEA) Conference on Teaching and Research in Economic Education (CTREE), Atlanta, GA, USA.

## APPENDIX 1

**The Visual Big Picture Of Macroeconomics Proposed By Moosavian (2016a)**

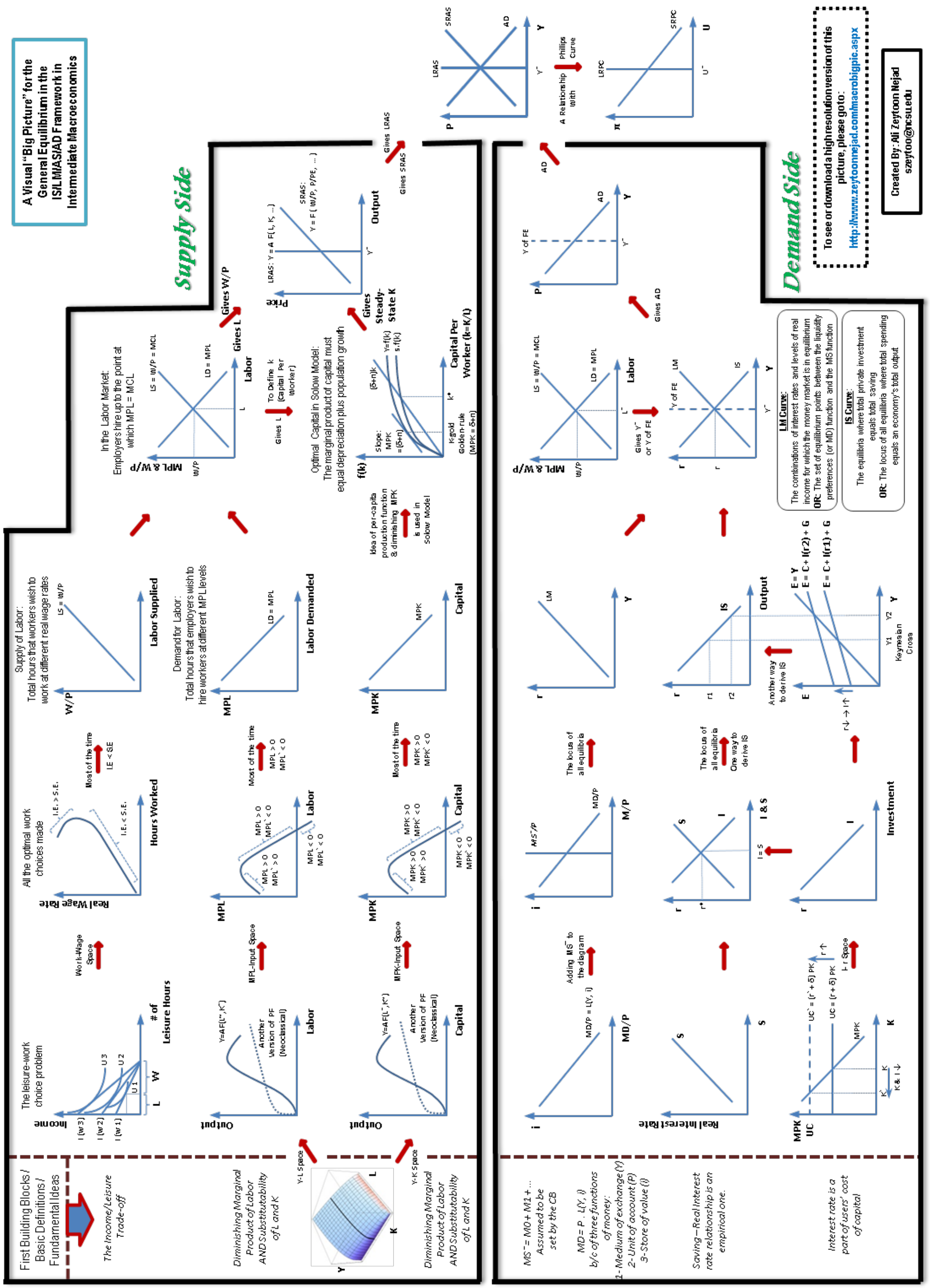

## APPENDIX 2

**The visual "big picture" of intermediate macroeconomics in terms of different markets and economics schools of thought**

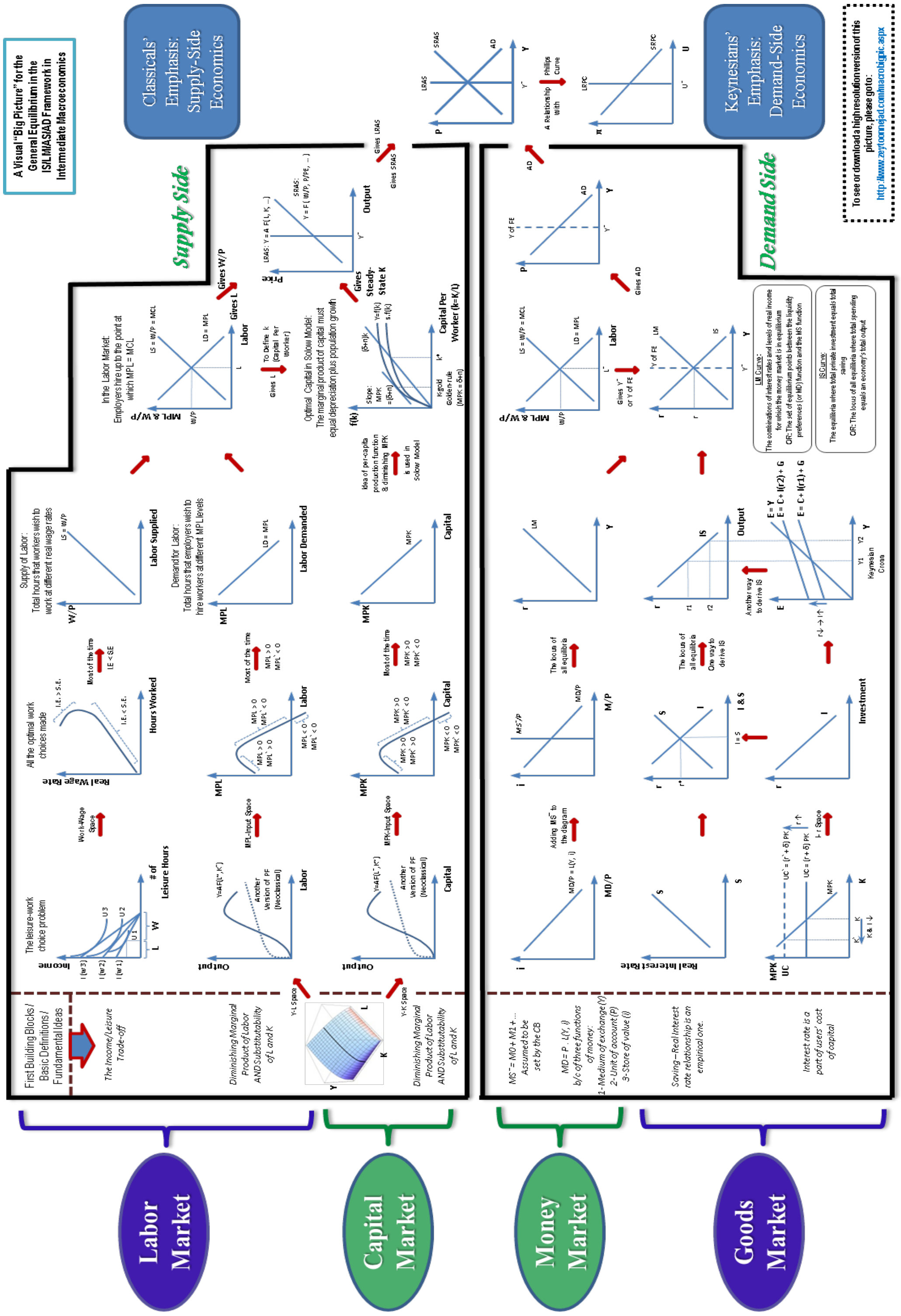

## APPENDIX 3

### The List of the Names of the Diagrams and Models

1. The Leisure-Work Choice Problem
2. Individual Labor Supply Curve
3. Labor Supply Diagram
4. Two-dimensional Production Function Diagram (Y-L Space)
5. Marginal Product of Labor (MPL) Diagram
6. Labor Demand Diagram
7. Labor Market Equilibrium Diagram
8. Three-dimensional Production Function Diagram (Y-L-K Space)
9. Two-dimensional Production Function Diagram (Y-K Space)
10. Marginal Product of Capital (MPK) Diagram
11. Capital Demand Diagram
12. Solow Model
13. Aggregate Supply (AS) Diagram
14. A Diagram for General Equilibrium in the Macroeconomy
15. Money Demand Diagram
16. Money Market Equilibrium Diagram (Money Supply and Demand)
17. LM Diagram (Liquidity-Money Diagram)
18. Labor Market Equilibrium Diagram
19. Aggregate Demand (AD) Diagram
20. Phillips Curve
21. Saving vs. Interest Rate Diagram
22. National Saving and Investment Model (aka "Classical Cross" Model)
23. IS Diagram (Investment=Saving Diagram)
24. IS-LM Model
25. User Cost of Capital Model
26. Investment vs. Interest Rate Diagram
27. Aggregate Expenditure Line (aka "Keynesian Cross" Model)

## APPENDIX 4

### Symbols and Notations

| | | |
|---|---|---|
| ‾ | : | Fixed |
| ~ | : | Changing |
| I | : | Income |
| I(w 1) | : | Income at wage rate 1 |
| U | : | Utility level OR Utility Indifference Curve |
| L | : | Leisure hours |
| W | : | Hours Worked |
| W | : | Nominal wage rate |
| w | : | Real wage rate OR W/P |
| P | : | Price level |
| I.E | : | Income effect |
| S.E | : | Substitution effect |
| L | : | Labor supplied OR Labor hours worked |
| LS | : | Labor supplied (Supply of labor) |
| LD | : | Labor demand (Demand for labor) |
| MCL | : | Marginal cost of labor |
| MPL | : | Marginal product of labor |
| Y | : | Output (Income) |
| A | : | Technology level OR Total Factor Productivity (TFP) level |
| PF | : | Production function |
| MPL’ | : | Derivative of marginal product of labor with respect to L |
| MPK | : | Marginal product of capital |
| MPK’ | : | Derivative of marginal product of capital with respect to K |
| F(.) OR f(.) | : | Function of |
| δ | : | Depreciation rate |
| LRAS | : | Long-run aggregate supply |
| SRAS | : | Short-run aggregate supply |
| AS | : | Aggregate supply |
| AD | : | Aggregate demand |
| FE | : | Full employment |
| Y‾ of Y of FE | : | Output level at full employment |
| PE | : | Price expectation |
| MS‾ | : | Money supply |
| M0 | : | Sum of currency in circulation (notes and coins) plus banks' reserves with the central bank |
| M1 | : | Currency in circulation plus current (checking) accounts plus deposit accounts transferable by checks |
| i | : | Nominal interest rate |
| r | : | Real interest rate |
| MD | : | Money demand (Demand for money) |
| LM | : | Liquidity-Money equilibrium curve |
| L(Y, i) | : | Liquidity function |
| S | : | National saving |
| I | : | National investment |
| IS | : | Investment-Saving curve |
| K | : | Capital stock |
| UC | : | User cost of capital |
| E | : | Expenditures |
| G | : | Government Expenditures |
| I(r1) | : | Investments made at the interest rate “r1” |
| C | : | Consumption |
| π | : | Inflation rate |
| U | : | Unemployment rate |
| U¯ | : | The natural rate of unemployment |
| LRPC | : | Long-run Philips curve |
| SRPC | : | Short-run Philips curve |

## APPENDIX 5

**An example of the Interactive Graphic Syllabus for the course of intermediate macroeconomics** (The graphic part of the paper version of the Interactive Graphic Syllabus)

Please See The Graphic Part On The Next Page …

**An Example Of Text Related To The Graphic Part**

**Graphic Syllabus of the Course:** In order for you to have a better understanding of the numerous, various topics that are planned to be discussed during this course as we move forward, please have a look at the graphic presented in the next page that illustrates a visual big picture showing how macroeconomic general equilibrium is formed, and also how this course is designed to go ahead. An appendix attached to this syllabus describes the symbols and notations used in this big picture, which are the most common symbols and notations employed in macroeconomics, which you need to be aware of. The web-page is accessible at the URL http://zeytoonnejad.com/macrobigpic.aspx and is still under minor modification. This web-page has been designed such that it works properly on almost all popular internet browsers, such as Firefox, Internet Explorer, Chrome, and Safari.

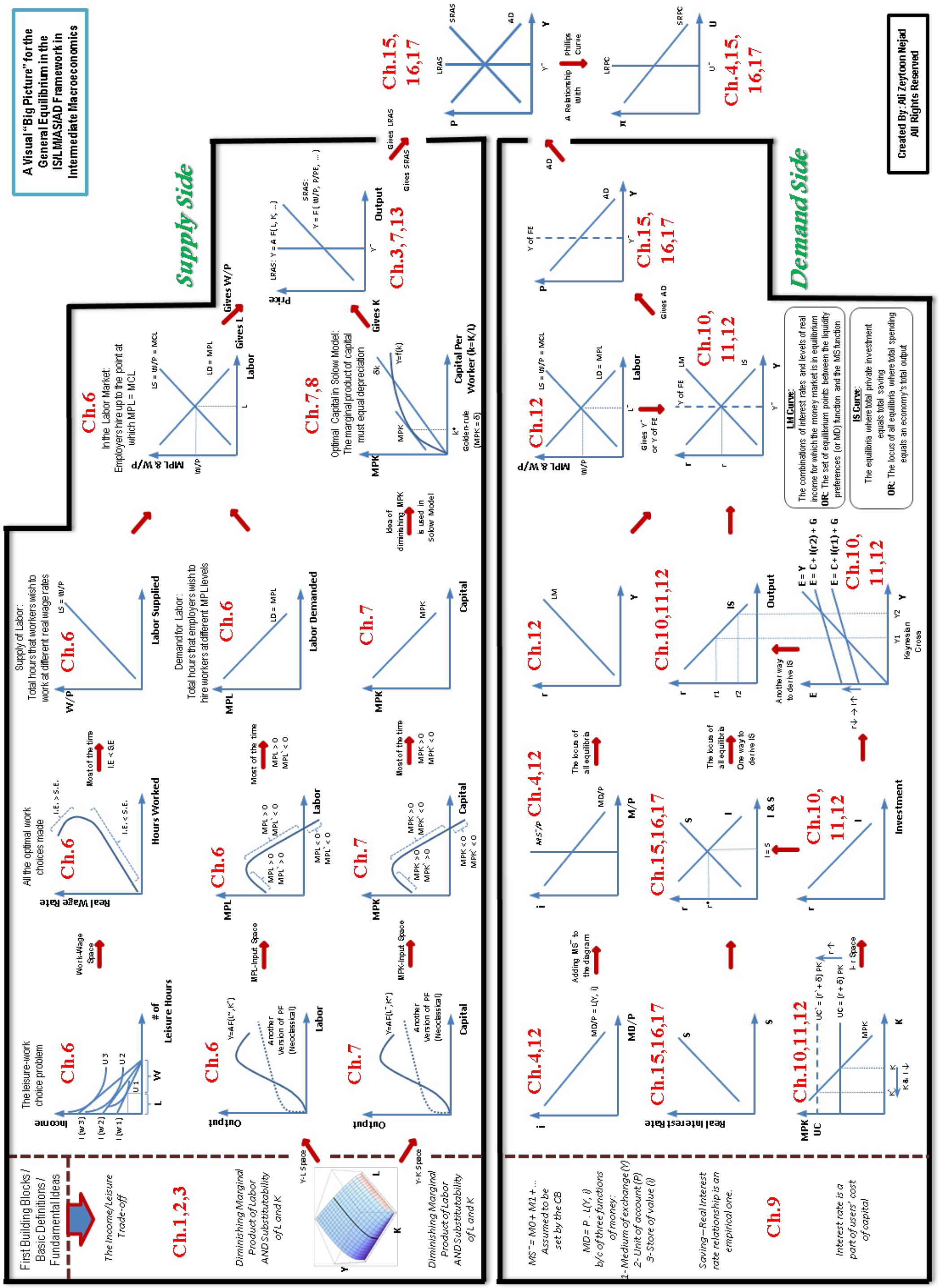

*Caveat:* You should already know about a few parts of this "big picture"; however, I will post some refreshers on the course website well ahead of time so that you can take advantage of them to catch up on the necessary background knowledge before we start talking about the materials associated with those parts. You can also use your own old lecture notes, textbooks, etc. if you prefer to refresh and reactivate the background knowledge you need to have before proceeding with the main subject matters of this course.

**APPENDIX 6**

**Information on computer software for designing the graphic part of an interactive graphic syllabus**

| Product | Price, Education Discount | Operating System(s) | Company Information | |
|---|---|---|---|---|
| | | | Name | Telephone Web Site |
| ConceptDraw MINDMAP 4 | $99 Yes | Windows Macintosh | Computer Systems Odessa, LLC | 1-877-441-1150 www.conceptdraw.com |
| EDGE Diagrammer | $59.95 Yes | Windows | Pacestar Software | 1-408-893-6046 www.pacestar.com |
| Inspiration 8 | $69 No | Windows Macintosh | Inspiration Software, Inc. | 1-800-877-4292 www.inspiration.com |
| MindGenius Education V2 | $107 Yes | Windows | Gael. Ltd. | +44 (0) 355-247766 |
| MindManager Basic 6, Pro 6 | $115, $149 Yes | Windows | Mindjet | 1-877-646-3538 |
| MindMapper Professional 5 | 179.95 Yes | Windows Macintosh | SimTech Systems, Inc. | 1-972-436-0863 or 1-877-883-6505 www.mindmapperusa.com/products.htm |
| VisiMap Professional 4 | £19 Yes | Windows | CoCo Systems Ltd. | +44 7971-321586 or 1-800-884-0489 (orders) www.coco.co.uk |
| Visual Mind 8 Basic, Business | $99, $149 No | Windows | Mind Technologies AS | +47 3285 5455 www.visual-mind.com |

**Source:** Nilson (2009)

**NOTES**